\documentclass[modern]{aastex62}

\usepackage{placeins}

\graphicspath{{./}{figures/}}

\shorttitle{LD2 Observations}
\shortauthors{Kareta et al.}

\begin{document}

\title{Contemporaneous Multi-Wavelength and Precovery Observations \\
of Active Centaur P/2019 LD2 (ATLAS)}

\correspondingauthor{Theodore Kareta}
\email{tkareta@lpl.arizona.edu}

\author[0000-0003-1008-7499]{Theodore Kareta}
\affil{Lunar and Planetary Laboratory, University of Arizona, Tucson, AZ, USA}

\author[0000-0002-0004-7381]{Laura M. Woodney}
\affil{Department of Physics, California State University San Bernardino, San Bernardino, CA, USA}

\author[0000-0003-1800-8521]{Charles Schambeau}
\affil{Florida Space Institute, University of Central Florida, Orlando, FL, USA}
\affil{Department of Physics, University of Central Florida, Orlando, FL, USA}

\author[0000-0003-1156-9721]{Yanga Fernandez}
\affil{Department of Physics, University of Central Florida, Orlando, FL, USA}
\affil{Florida Space Institute, University of Central Florida, Orlando, FL, USA}

\author[0000-0002-2014-8227]{Olga Harrington Pinto}
\affil{Department of Physics, University of Central Florida, Orlando, FL, USA}

\author{Kacper Wierzchos}
\affil{Lunar and Planetary Laboratory, University of Arizona, Tucson, AZ, USA}

\author[0000-0003-4659-8653]{M. Womack}
\affil{Florida Space Institute, University of Central Florida, Orlando, FL, USA}
\affil{Department of Physics, University of Central Florida, Orlando, FL, USA}
\affil{National Science Foundation, Alexandria, VA USA}

\author{S.J. Bus}
\affil{Institute for Astronomy, University of Hawaii, Honolulu, HI, USA}

\author{Jordan Steckloff}
\affil{Planetary Science Institute, Tucson, AZ, USA}
\affil{University of Texas at Austin, Austin, TX, USA}

\author[0000-0001-5678-5044]{Gal Sarid}
\affil{SETI Institute, Mountain View, CA, USA}

\author[0000-0001-8736-236X]{Kathryn Volk}
\affil{Lunar and Planetary Laboratory, University of Arizona, Tucson, AZ, USA}

\author{Walter M. Harris}
\affil{Lunar and Planetary Laboratory, University of Arizona, Tucson, AZ, USA}

\author{Vishnu Reddy}
\affil{Lunar and Planetary Laboratory, University of Arizona, Tucson, AZ, USA}

\begin{abstract}

Gateway Centaur and Jupiter co-orbital P/2019 LD2 (ATLAS) \citep{Sarid:2019} provides the first opportunity to observe the migration of a Solar System small body from a Centaur orbit to a Jupiter Family Comet (JFC) four decades from now \citep{Kareta:2020, 2020arXiv200713945H}. The Gateway transition region is beyond where water ice can power cometary activity, and coma production there is as poorly understood as in all Centaurs. We present contemporaneous multi-wavelength observations of LD2 from 2020 July 2-4: Gemini-North visible imaging, NASA IRTF near-infrared spectroscopy, and ARO SMT millimeter-wavelength spectroscopy. Precovery DECam images limit the nucleus's effective radius to $\leq$ 1.2 km and no large outbursts were seen in archival Catalina Sky Survey observations. LD2's coma has $g' - r' = 0.70 \pm 0.07$, $r' - i' = 0.26 \pm 0.07$, a dust production rate of $\sim 10-20$ kg/s, and an outflow velocity between $ v \sim 0.6-3.3$ m/s. We did not detect CO towards LD2 on 2020 Jul 2-3, with a 3-$\sigma$ upper limit of $Q(CO) < 4.4\times10^{27}$ mol s$^{-1}$ ($\lessapprox$ 200 kg s$^{-1}$). Near-infrared spectra show evidence for water ice at the 1-10\% level depending on grain size. Spatial  profiles and archival data are consistent with sustained activity. The evidence supports the hypothesis that LD2 is a typical small Centaur that will become a typical JFC, and thus it is critical to understanding the transition between these two populations. Finally, we discuss potential strategies for a community-wide, long baseline monitoring effort.

\end{abstract}


\section{Introduction} \label{sec:intro}

The active outer Solar System body P/2019 LD2 (ATLAS) was discovered by the ATLAS project on June 10, 2019, with already published precovery observations taken at least as early as May 21, 2018 with Pan-STARRS 1 and December 01, 2017 with Zwicky Transient Facility (CBET 4780, MPEC 2020-K134). Although LD2 was first classified as member of the Jupiter Trojan asteroid population due to it having a semimajor axis very near that of Jupiter, detailed dynamical studies \citep{Kareta:2020, 2020arXiv200713945H, Steckloff:2020} revealed that LD2's current orbit has a different origin. LD2 is actually a persistently active member of the Centaurs, a dynamically unstable population in transition from the trans-Neptunian region into the Jupiter Family of Comets (JFCs) in the inner Solar System; only in 2017 was LD2 scattered by Jupiter into its current orbit, which has a perihelion distance of $q=4.57 AU$, a Tisserand parameter with respect to Jupiter of $T_{J}=2.941$, and is co-orbital with Jupiter. This contrasts with the Jupiter Trojan asteroids which were emplaced during the era of Giant Planet migration $\sim$ 4 Gyr ago. LD2 is currently near the “Dynamical Gateway” orbital region \citep{Steckloff:2020}, which facilitates most of the dynamical transitions between the Centaur and JFC populations \citep[see][]{Sarid:2019}. These dynamical studies also found that LD2 is unlikely to have spent significant time in the inner Solar System where water ice sublimes vigorously and drives activity. This strongly suggests that LD2 is a pristine object that has only been affected by the thermal environment present in the Centaur region \citep{Steckloff:2020} and thus is in a thermal state immediately prior to that of the JFCs evolutionarily. 

Perhaps most exciting is the discovery that LD2 is very likely ($>98\%$) to become a JFC after a close encounter with Jupiter in 2063 \citep{Kareta:2020, 2020arXiv200713945H}. This is the first opportunity to observe how a pristine Centaur responds to the changing thermal environment as it completes its dynamical evolution \citep{Steckloff:2020}. Thus far, all JFCs have been observed after this transition has already happened, such as 81P/Wild 2 \citep[see][]{2004Sci...304.1764B, 2006Sci...314.1711B,Brownlee2014}, which was found shortly after it entered the inner Solar System in the 1970s. The orbital transition from middle Solar System Centaurs to the more extreme thermal environment of the inner Solar System JFCs results in a large change in a comet's thermophysical state, facilitating changing and increasing outgassing patterns from a variety of volatiles and greatly changing the evolution of the solid body. Gaseous emission has been observed in some Centaurs and is dominated by CO and CO$_2$  \citep{bus91, cochransw1copcn91, senayjewitt94, womacksternchiron99, wierzchosecheclus2017}. Inwards of $\sim3$ AU, carbon monoxide (CO) and/or carbon dioxide (CO$_2$) sublimation is typically no longer the dominant driver of cometary activity, but rather is dominated by crystalline water ice sublimation \citep{Womack2017DistantReview}. There are exceptions, such as hyperactive comet 103P/Hartley 2, whose activity is dominated by CO$_2$ sublimation \citep{AHearn:2011}, likely due to a recent mass-wasting event that exposed this supervolatile ice \citep{Steckloff:2016}. These orbital transitions thus mark not only where an object becomes a JFC from a dynamical standpoint, but also where the character of its activity changes and it starts acting like one. 
LD2's place in the Gateway and imminent inwards transition to the Jupiter Family is thus a critical opportunity to study these processes in detail to improve our understanding of the evolution of JFCs in general.

It is therefore critical to establish baseline measurements of LD2's activity to facilitate future comparisons of LD2's activity as it responds to the changing thermal environment characteristic of this Centaur-to-JFC transition early in the process of this orbital change. This includes but is not limited to characterizing its dust coma in size, brightness, morphology and composition as well as searches for volatile outgassing or transient behaviors. While we know at the current moment that LD2 has been active for at least a few years (its most recent perihelion date was April 11, 2020), it is unclear whether this activity level has varied considerably or has remained steady. If LD2's activity had modulated greatly between individual attempts to characterize it, inferences from one epoch might lead to misunderstandings about the general activity level of the object. However, high-quality characterization in the short term without a proper understanding of its long-baseline activity is equally prone to misunderstandings.

In this paper, we present a series of observations and archival data searches to address these obstacles and properly characterize this compelling object. In Section 2, we present and analyze serendipitous observations taken prior to the official report of activity to better constrain its general activity state and nuclear size. We then present simultaneous observations taken at visible, near-infrared, and millimeter wavelengths with the Gemini North telescope, NASA Infrared Telescope Facility (IRTF), and Arizona Radio Observatory 10-m Submillimeter Telescope (SMT) in early July 2020 to characterize LD2. In Section 3, we apply models to estimate the size of the nucleus of LD2, the general activity level and dust properties, as well as the water ice content in the inner coma. Lastly, in Section 4 we analyze and compare all of our results to put LD2 into context as an active Centaur and future JFC.

\section{Observations and Data Reduction} \label{sec:obs}
In this section, we detail the UT July 2020 multi-wavelength observations of active Centaur LD2, precovery imaging data obtained from the Dark Energy Survey and Catalina Sky Survey archives, and methods by which each dataset was reduced. Planning for these observations began in late May and early June 2020 when the importance of LD2 became clear. Proposals for Directors Discretionary Time (DDT) were sent in June to Gemini North, the NASA IRTF, and the Arizona Radio Observatory SMT to coincide on the July 2-4 observing time frame. The DECam archive was searched in the late Summer of 2020, and the CSS archive was queried and analyzed in October and November 2020.

\subsection{Optical Observations}
Multi-epoch visible imaging data of LD2 was analyzed from the telescopes listed in Table \ref{tab:vis_phot} in order to monitor its dust coma activity. All imaging data from Gemini North and DECam underwent photometric calibration using reported Pan-STARRS magnitudes of image field stars \citep{maginer_2013}, while the reduction of the Catalina dataset is described in section 2.1.2.

\begin{deluxetable*}{cccccccccc}[b!]
\tablecaption{Optical Imaging Observing Geometry\label{tab:vis_phot}}
\tablecolumns{10}
\tablewidth{0pt}
\tablehead{
\colhead{UT Date} &
\colhead{UTC Time\tablenotemark{a}} &
\colhead{$R_H$} &
\colhead{$\Delta$} &
\colhead{$\alpha$} &
\colhead{Seeing} &
\colhead{Tel./Inst./Filter} &
\colhead{Total Exp. Time} &
\colhead{Airmass\tablenotemark{b}} &
\colhead{Active}\\      
\colhead{(YYYY-MM-DD)} & 
\colhead{(d)} &
\colhead{(au)} &
\colhead{(au)} &
\colhead{$(^{\circ})$} &
\colhead{(arcsec)} & 
\colhead{} & 
\colhead{(seconds)} &
\colhead{} &
\colhead{(Y/N/?)}
}
\startdata
2017-03-06 & 07:32:44 & 5.52 & 4.69 & 6.18  & $\sim 1\arcsec.0$ & Blanco/DECam/r$'$ & 43 & 1.09 & ?  \\
2018-08-10 & 00:00:22 & 4.90 & 4.78 & 11.92 & $\sim 0\arcsec.8$ & Blanco/DECam/g$'$ & 68 & 1.18 & Y  \\
2020-07-03 & 13:33:26 & 4.58 & 3.76 & 8.24  & $\sim 0\arcsec.5$ & Gemini-N/GMOS/r$'$ & 450 & 1.11 & Y  \\
2020-07-03 & 13:45:31 & 4.58 & 3.76 & 8.24  & $\sim 0\arcsec.5$ & Gemini-N/GMOS/g$'$ & 300 & 1.13 & Y  \\
2020-07-03 & 13:53:33 & 4.58 & 3.76 & 8.24  & $\sim 0\arcsec.5$ & Gemini-N/GMOS/i$'$ & 300 & 1.14 & Y  \\
\enddata
\tablenotetext{a}{ UTC at start of sequence or image exposure. See text for magnitudes and upper limits.}
\tablenotetext{b}{Mean airmass of LD2 during exposure.}
\end{deluxetable*}
\FloatBarrier

\begin{deluxetable*}{ccccccccc}[b!]
\tablecaption{CSS Optical observations\label{tab:CSS_obs}}
\tablecolumns{8}
\tablewidth{0pt}
\tablehead{
\colhead{UT Date} &
\colhead{UTC Time\tablenotemark{a}} &
\colhead{$R_H$} &
\colhead{$\Delta$} &
\colhead{$\alpha$} &
\colhead{Instrument} & 
\colhead{Total Exp. Time} &
\colhead{Airmass} &
\colhead{Magnitude\tablenotemark{b}} \\
\colhead{(YYYY-MM-DD)} & 
\colhead{(d)} &
\colhead{(au)} &
\colhead{(au)} &
\colhead{$(^{\circ})$} &
\colhead{} & 
\colhead{(seconds)} & 
}
\startdata
2016-01-31 & 11:02:24 & 5.65 & 4.85 & 6.31 & 0.7m Schmidt   & 120 & 1.23 & $>$20.1  \\
2016-02-15 & 08:24:00 & 5.64 & 4.71 & 3.94 & 0.7m Schmidt   & 120 & 1.21 & $>$20.5  \\
2016-03-03 & 07:55:12 & 5.63 & 4.64 & 1.29 & 0.7m Schmidt   & 120 & 1.19 & $>$20.7 \\
2017-04-05 & 07:55:12 & 5.47 & 4.48 & 0.94 & 1.5m   & 120 & 1.25 & $>$21.5 \\
2017-04-18 & 06:28:48 & 5.45 & 4.47 & 1.99 & 1.5m    & 120 & 1.24 & $>$21.7 \\
2018-03-21 & 11:45:36 & 5.05 & 4.44 & 9.54 & 1.5m   & 120 & 1.37 & $>$21.1 \\
2018-05-14 & 06:57:35 & 4.99  & 3.99 & 2.27 & 1.5m    & 120 & 1.31 & $\geq$21.9  \\
2018-06-18 & 05:45:37 & 4.95 & 4.13 & 7.57 & 1.5m   & 120 & 1.32 & $\geq$21.5 \\
2020-04-19 & 11:16:48 & 4.57 & 4.77 & 12.07 & 1.5m   & 120 & 2.13 & $18.57\pm0.08$ \\
2020-04-21 & 11:31:12 & 4.57 & 4.75 & 12.18 & 0.7m Schmidt    & 120 & 1.89 & $18.54\pm0.36$  \\
2020-04-27 & 11:16:51 & 4.57 & 4.66 & 12.64 & 0.7m Schmidt    & 120 & 1.83 & $18.48\pm0.26$  \\
2020-05-22 & 09:07:12 & 4.57 & 4.54 & 12.68 & 0.7m Schmidt   & 120 & 2.23 & $18.71\pm0.15$  \\
2020-05-24 & 10:33:36 & 4.57 & 4.26 & 12.50 & 1.5m    & 120 & 1.47 & $18.74\pm0.05$  \\
2020-07-18 & 08:09:33 & 4.58 & 3.65 & 5.58 & 0.7m Schmidt    & 120 & 1.28 & $17.69\pm0.04$  \\
2020-07-30 & 06:57:36 & 4.58 & 3.60 & 3.30 & 0.7m Schmidt    & 120 & 1.31  & $17.62\pm0.40$ \\
2020-08-19 & 06:00:03 & 4.59 & 3.60 & 3.19 & 1.5m   & 120 & 1.30 & $17.73\pm0.05$  \\
2020-08-26 & 06:28:29 & 4.59 & 3.63 & 4.47 & 0.7m Schmidt   & 90 & 1.31 & $17.77\pm0.15$  \\
2020-09-07 & 05:16:48 & 4.59 & 3.71 & 6.77 & 0.7m Schmidt    & 120 & 1.31 & $18.05\pm0.05$ \\
2020-09-11 & 02:38:24 & 4.60 & 3.75 & 7.49  & 1.5m    & 120 & 1.61 & $17.79\pm0.05$  \\
2020-09-24 & 02:52:53 & 4.60 & 3.89 & 9.55 & 1.5m   & 120 & 1.39 & $17.85\pm0.09$  \\
2020-10-06 & 02:38:21 & 4.60 & 4.04 & 10.98 & 0.7m Schmidt   & 120 & 1.36 & $18.16\pm0.25$ \\
2020-10-19 & 02:52:45 & 4.61 & 4.23 & 11.96 & 1.5m    & 120 & 1.36 & $18.24\pm0.04$  \\
2020-10-21 & 23:16:50 & 4.61 & 4.26 & 12.06 & 0.7m Schmidt   & 120 & 1.97 & $18.32\pm0.19$  \\
2020-10-29 & 03:36:00 & 4.61 & 4.38 & 12.32 & 1.5m    & 120 & 1.55 & $18.53\pm0.12$  \\
2020-11-05 & 01:40:47 & 4.62 & 4.51 & 12.38 & 0.7m Schmidt    & 120 & 1.37 & $18.05\pm0.16$ \\
\enddata
\tablenotetext{a}{ UTC at the middle of sequence or image exposure.}
\tablenotetext{b}{All images taken with a clear filter, and the details of the photometric calibration are described in the text.}
\end{deluxetable*}
\FloatBarrier

\subsubsection{Precovery DECam Imaging Data}

We used the Canadian Astronomy Data Centre's Solar System Object Image Search \citep{gwyn_2012_pasp} to search the Dark Energy Survey (DES; \citealt{abbott_2018}) archive for serendipitous observations of LD2 from the Dark Energy Camera (DECam; \citealt{decam_2015_aj}) installed on the Cerro Tololo Inter-American Observatory's Blanco 4-meter telescope. The DECam instrument consists of 62 individual 2k$\times$4k CCDs for imaging in a hexagonal arrangement on the focal plane. This results in a 2.2 degree diameter field of view and 0$''$.263 per pixel plate scale. The combination of a relatively large field of view and the deep imaging afforded by the Blanco telescope results in a treasure trove of serendipitously imaged Solar System small bodies using DECam.

In total, our DES search revealed six epochs of precovery LD2 imaging data ranging from UT 2014-12-30 to 2018-08-10. DECam's archived multi-extension FITS images have undergone basic reductions including bias subtraction, flat field correction, and fitting of a World Coordinate System as described in \cite{morganson_2018}. We also applied our Python-based DECam image processing pipeline to individual images to: (1) identify the FITS extension containing coverage of LD2 based on its JPL Horizons ephemeris coordinates, (2) save the identified individual CCD data and header metadata as a new FITS image file, (3) perform cosmic ray removal utilizing the LACosmic technique \citep{van_dokkum_2001} as implemented in ccdpro \citep{matt_craig_2017}, and (4) perform photometric calibration based on Pan-STARRS field stars.

LD2 was not detected in the DECam images prior to UT 2017-05-15 within 10$''$ of its ephemeris position (1-sigma ephemeris uncertainty $\pm$ 0$''$.2) as generated by the orbit in MPEC 2020-O07. LD2 was detected in a VR image on UT 2018-05-09 and g$'$ image on UT 2018-08-10 where predicted by the ephemeris within errors. These data were used to assess when LD2's coma activity began and for estimates of its nucleus size. 

\subsubsection{Precovery Data from the Catalina Sky Survey}

Precovery observations of LD2 were found in data from the 0.7m Schmidt and the 1.5m reflector of the Catalina Sky Survey (CSS). Both telescopes are located in the Santa Catalina Mountains in southern Arizona. The 0.7m Schmidt (MPC code 703) is located on Mt. Bigelow, and the 1.5m (MPC code G96) is located on Mt. Lemmon. The same 10k$\times$10k CCD camera is used in both telescopes, where it is mounted on prime focus. The fields of view and platescales at the 0.7m Schmidt and 1.5m are 19.4 deg.$^{2}$ with 3$''/$pixel, and 5 deg.$^{2}$ with 1.44$''/$pixel respectively. During survey operations, both telescopes use a nominal exposure time of 30s with 4 visits per field. All exposures were unfiltered and a photometric calibration was applied with respect to GAIA DR2 \citep{Gaia_2018}. This results in a limiting unfiltered Gaia G-magnitude of $\sim$ 19.5 at the 0.7m Schmidt and $\sim$ 21.8 at the 1.5m, depending on atmospheric conditions.
A search for precovery observations in the CSS image archive was carried out using a proprietary CSS pipeline. This search resulted in the positive identification of LD2 in 10 epochs between UT 2020-04-21 and UT 2020-11-05 in data from the 0.7m Schmidt and in 9 epochs between UT 2018-05-14 and UT 2020-10-29 in data from the 1.5m (see Table \ref{tab:CSS_obs}). LD2 was not detected in 0.7m Schmidt images obtained between UT 2016-01-31 and UT 2018-05-14 within 60'of its ephemeris position calculated from the available 920-day arc astrometry contained in the MPC Orbits/Observations database. Two positive but marginal detections of LD2 were found in data from the 1.5m telescope on UT 2018-05-14 and UT 2018-06-18. In those detections, the object was near or at the nominal limiting magnitude of the 1.5m telescope and hence it is not possible to ascertain whether the object was active or not on those epochs based on the two images.

\subsubsection{Gemini-N Observations}\label{ss:gemini}
\begin{figure}[ht!]
 \centering
\includegraphics[width=0.975\textwidth]{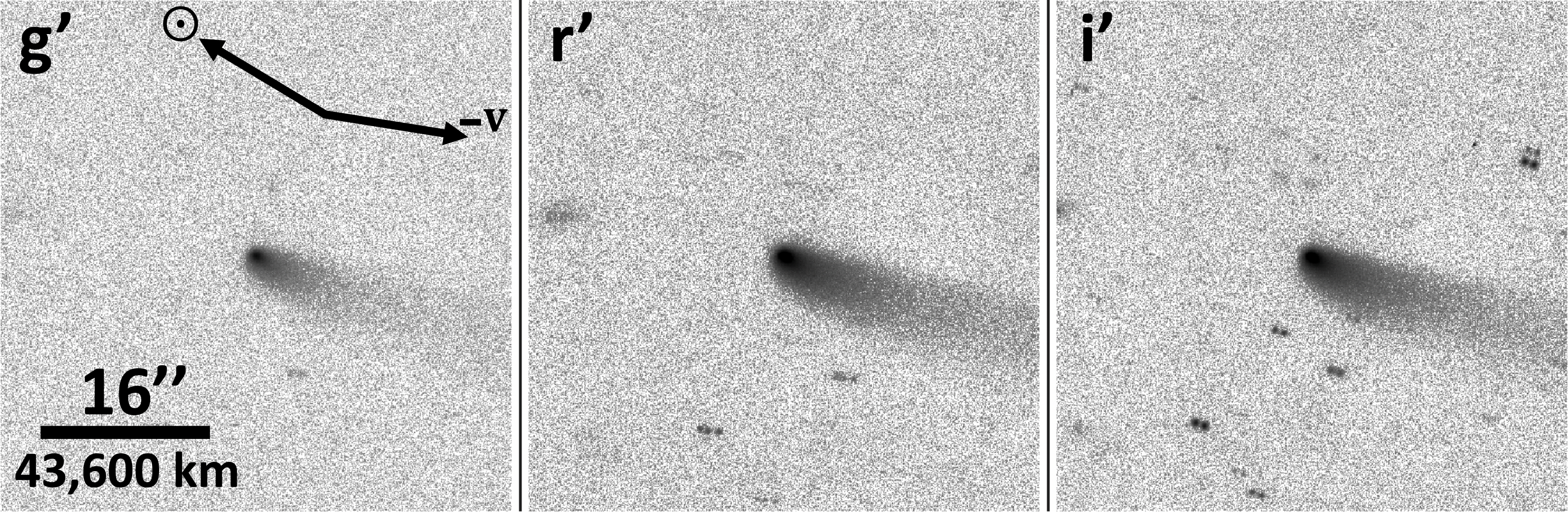}
\caption{Panels display the July 3, 2020 GMOS observations using an inverted gray logarithmic scale. Images are the result of stacking individual 150 second exposure to increase the detection of low-level surface brightness of the dust coma. Total exposure times respectively are: g$'$ (300 seconds), r$'$ (450 seconds), and i$'$ (300 seconds). The orientations and scales of the panels are the same for each, having equatorial north up and east to the left. Arrows indicate the position angles of the skyplane projected LD2-Sun vector and negative heliocentric velocity vector. We note that due to the very low phase angle of these images, most of LD2's dust tail is directly away from the observer (e.g. into the page) and thus not clearly discerned.}
\label{fig:gemini_images}
\end{figure}

Observations of LD2 were acquired from the Gemini North observatory (Program ID: GN-2020A-DD-202; PI C. Schambeau) using the Gemini Multi-Object Spectrograph (GMOS) in imaging mode on UT 2020-07-03. No detector binning was used, maintaining the $\sim$ 0$''$.08/pixel spatial sampling. Images were acquired in the GMOS g$'$ ($475 nm$), r$'$ ($620 nm$) and i$'$ ($720 nm$) filters using individual exposures of 150 seconds. Figure \ref{fig:gemini_images} displays examples of the GMOS images centred and cropped around LD2. Table \ref{tab:vis_phot} provides details of the observing circumstances and photometric measurements. Sky conditions during UT 2020-07-03 image acquisition were stable and photometric with an average seeing of 0$''$.5.

All Gemini GMOS data were reduced, including bias subtraction, flat field application, and mosaicking, using the Gemini IRAF software \citep{gemini_iraf_2016}. Additionally, cosmic ray removal and Pan-STARRS based calibration were applied to individual image frames following a similar procedure as described for the DECam data.

During the observations a malfunction of the OIWFS guide camera resulted in the images being acquired while the telescope tracked at the sidereal rates. Due to LD2's slow apparent motion on the projected skyplane ($<12''$/hour) and the requested 150 second individual exposures, there is minimal blurring of the central peak in each individual image.

\subsection{Near-Infrared Observations}
Observations of P/2019 LD2 (ATLAS) were obtained from the NASA IRTF on 2020 July 2 and 3 (PI: L. Woodney) via DDT. We used the SpeX instrument \citep{rayner_spex:_2003} in the low-resolution `prism' mode ($R \sim{100}$), providing an effective wavelength range of $\sim0.7\mu{m}$ to $\sim2.5\mu{m}$. As LD2 was expected to be a dim ($m_{V}\sim{18.6}$) and extended target, we used the MORIS camera \citep{2011PASP..123..461G} for guiding as well as broadband imaging. SpeX's slit was chosen to be $0.8\arcsec$  wide and $15\arcsec$ long, and MORIS's LPR600 filter was used to maximize signal to noise in short exposures for optimum guiding; both choices were based on experience observing similar targets with the same instruments. The dichroic filter which split the light between the two instruments was set to $0.8\mu{m}$, resulting in slightly lower signal at shorter wavelengths in the science data. The instrument rotator was set to be North-South (as opposed to the parallactic angle) to conserve observing time, but we note that atmospheric dispersion is minimal at low airmass, in the near-infrared, and for extended targets. Under typical conditions at the IRTF, \cite{rayner_coolstars_2009} found a $\sim 0.3\arcsec$ dispersion between $0.8 \mu m$ and $2.4 \mu m$ at our largest airmass ($1.25$), which is significantly smaller than the slit width, seeing, and most importantly the size of LD2's coma. As a result, we do not believe atmospheric diffraction to be an issue for these observations, certainly so beyond $1.0 \mu m$.

\begin{deluxetable*}{cccccCrlcc}[b!]
\tablecaption{Summary of Infrared Observational Data\label{tab:observations}}
\tablecolumns{9}
\tablenum{3}
\tablewidth{0pt}
\tablehead{
\colhead{UT Date\tablenotemark{a}} &
\colhead{UTC Time\tablenotemark{a}} &
\colhead{$R_H$} &
\colhead{$\Delta$} &
\colhead{$\alpha$} &
\colhead{Seeing} & 
\colhead{Inst./Setting\tablenotemark{b}} & 
\colhead{Total Time}\\      
\colhead{(YYYY-mm-dd)} & 
\colhead{(d)} &
\colhead{(au)} &
\colhead{(au)} &
\colhead{$(^{\circ})$} &
\colhead{(arcsec)} & 
\colhead{} & 
\colhead{seconds}
}
\startdata
2020-07-02 & 10:30 & 4.58 & 3.77 & 8.44 & \sim 0.74\arcsec & SpeX/Prism & 8000.0  \\
2020-07-03 & 10:30 & 4.58 & 3.76 & 8.27 & \sim0.41\arcsec & SpeX/Prism  & 8400.0  \\
2020-07-04 & 14:25 & 4.58 & 3.75 & 8.10 & \sim0.5\arcsec & SpeX/Imaging & 1600.0  \\
\enddata
\tablenotetext{a}{ UTC at start of sequence or image exposure.}
\tablenotetext{b}{All observations taken simultaneous to imaging with the MORIS camera, though the primary use of those observations were for guiding.}
\end{deluxetable*}

On 2020 July 2 and 3, the total observational window subtended from 00:30 to 6:10 Hawaii time (10:30 to 16:10 UT), with LD2 transiting near the midpoint of the allotted time. Spectroscopic observations of the target were `book-ended' by observations of both local stars with Sun-like colors and a local A0 star to provide multiple ways to correct for telluric absorption in the observations of the target. This resulted in three sets of standard star observations before, in between, and after two blocks of observations of LD2. Each block was 20 exposures of 200 seconds each (in an ABBA nodding pattern), resulting in 8000 seconds ($\sim$2 hours and 12 minutes) of integration time on each night (July 3rd allowed for one more AB pair resulting in 8400 seconds total). Conditions were good on both nights, with average humidity of $15\%$ and $20\%$ and seeing of $0.74\arcsec$ and $0.41\arcsec$ on each night respectively, though we note that due to the fixed slit width and large size of LD2's coma this does not effect the retrieved resolution.

An additional shorter block of time on 2020 July 4 (PI: V. Reddy, Hawaii time 4:25 - 6:10) was used to acquire J-band photometry of LD2 using the SpeX instrument in imaging mode. A total of 8 exposures of 200 seconds each ($\sim$ 27 minutes of integration time) on LD2 was acquired, again book-ended by observations of a star with Sun-like colors nearby in the sky. These data were obtained both to directly compare with visible-wavelength imaging from Gemini, DECam, and CSS as well as an additional way to verify the calibration of the short wavelength end of the IRTF spectra that might have been affected by the dichroic cut-off.

The spectral observations were reduced independently using both the `spextool' package \citep{2004PASP..116..362C} as well as an independently written pipeline in IDL and the results were found to be identical. All science spectra were inspected individually for quality assurance (e.g., guiding issues, artifacts, etc) and were extracted with multiple apertures ($1.0\arcsec$, $2.0\arcsec$) to both maximize the signal-to-noise ratio and to look for spatial dependence of the reflectance spectrum retrieved as previously noted in similar observations of the active Centaur 174P/Echeclus \citep{2019AJ....158..255K}. The spectra of LD2 were then divided by spectra of the Sun-like stars (including a telluric correction) and averaged together. The MORIS (July 2, 3) and SpeX images (July 4) were reduced, stacked, and calibrated using standard methods for CCDs and near-infrared detectors in IDL and Python, respectively.

\subsection{Millimeter-wavelength Spectroscopy}

We used the Arizona Radio Observatory (ARO) 10-m Submillimeter Telescope (SMT) at Mt. Graham, Arizona to observe the CO J=2-1 transition at 230.537 GHz. We observed the comet during 2020 Jul 2 and 3 UT with heliocentric distances of $r$ = 4.58 au with geocentric distances of $\Delta$ = 3.78 au. Pointing and focus was updated every six 6-minute long scans on a planet or a bright radio-source. The accuracy of the pointing and tracking was $\sim$  2 arcsec RMS. We used beam-switching mode with a throw of 2 arcminutes for sky measurement. Due to the long, $\sim$ 3x10$^7$ s, lifetime of CO at 4.6 au, we estimate that $\sim$ 10\% of CO emission was subtracted with this procedure, which we take into account with our final CO production rate calculation.

We used the 1.3mm ALMA Band 6 dual polarization receiver where the chopper wheel method was used for setting the temperature scale T$_A^*$, with  T$_R$=T$_A^*$/{$\eta_b$}. System temperatures were an average of $T_{sys} \sim$ 980 K throughout the observing period with no interruptions from monsoon storms typical of that time of year.  Two spectral resolutions were available with the following filterbanks: 250 kHz/channel (500 channels in parallel) and 1000 kHz/channel (2048 channels in parallel). The bandwidth was 128 and 512 MHz respectively. Only the 250 kHz/channel filterbank was used as it provided a significantly better spectral resolution  of 0.33 km s$^{-1}$. We used the beam efficiency of $\eta_b$ = 0.74 provided by the observatory.

\section{Results} \label{sec:results}

\subsection{Nucleus Size Estimates and General Activity State}

Nucleus size estimates for LD2 place it in context with both the Centaur and JFCs nuclei size frequency distributions. LD2's discovery imaging by the ATLAS survey occurred during a period of ongoing coma activity. Since discovery, LD2 has maintained coma activity, confusing attempts to directly image its bare nucleus for robust size estimates. For this reason, we searched the DECam archive for serendipitous precovery imaging data for epochs with no detectable coma. 

\begin{figure}[ht!]
 \centering
\includegraphics[width=0.975\textwidth]{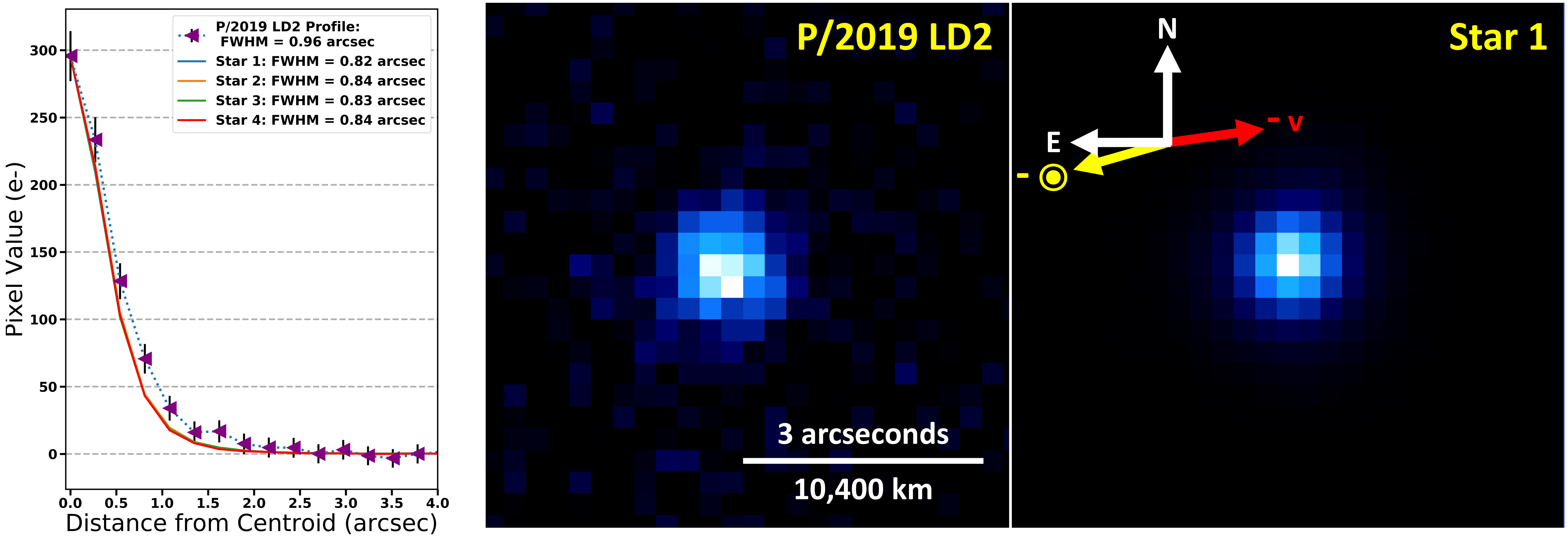}
\caption{Panels displaying the UT 2018 August DECam radial profile analysis. Left: LD2's radial profile width was compared to four field stars and found to be wider than the $\sim$ 0$''$.8 seeing limited stellar profiles. Middle: Cropped image of DECam detection of LD2. An asymmetric extended emission is faintly detected when compared to stars in the same DECam frame (Right image).}
\label{fig:decam_aug_2018}
\end{figure}

Figure \ref{fig:decam_aug_2018} shows DECam g$'$ imaging from UT  2018-08-10 and reveals LD2 as a condensed object with size 1$''$.0 (FWHM; compared to 0$''$.8 for nearby stars) located within 0$''$.1 of an ephemeris generated by the orbit in MPEC 2020-O07. Photometry of this source yields apparent magnitude g$'$ = 21.44 $\pm$ 0.05 and is consistent with Pan-STARRS1 reported photometry included in MPEC 2002-K134. Based on this photometry, an ersatz estimate for the nucleus's radius is 5.1 $\pm$ 0.1 km assuming a 5\% geometric albedo, or 3.4 $\pm$ 0.1 km assuming an 11.2\%  albedo (the average red-Centaur albedo reported by \citealt{Romanishin:2018}). 
However, this size is inconsistent with earlier DECam imaging data from 2017-03-06, where no source was detected within 10$''$ of the comet's expected ephemeris position (1-sigma ephemeris uncertainty $\pm$ 0$''$.2) as generated by the orbit in MPEC 2020-O07. Three-sigma upper-limit photometry on the March 6 r$'$ ($\leq$ 23.8) imaging data indicate that the upper limits on the nucleus's radius are 1.2 km assuming a 5\% albedo or 0.8 km assuming an 11.2\% albedo. This inconsistency may be the result of a compact coma present at the time of the 2018 Aug. 10 g$'$ imaging, thus confusing the nucleus-size estimate and/or the comet having an elongated nucleus. We consider these non-detections as useful constraints, and are consistent with observations of the most common sizes of JFC nuclei \citep{2004Icar..170..463M, bauer_2017, Snodgrass:2011, Fernandez:2013}.

While the sparse photometry presented here from DECam and CSS is inadequate to constrain hour-to-hour or day-to-day variations in brightness, they serve as a highly useful first-order look at how variable the overall state of the object is and as important context for the July 2020 observations. In the Catalina dataset, the variation in brightness visit-to-visit was almost always below $\sim0.2-0.3$ magnitudes, with the exception of between May and July 2020 where LD2 brightened by $\sim 1.0$ magnitude. LD2's brightness was stable then from 18 July 2020 through 28 August 2020. While it is possible that there are small or quickly dissipating outbursts occuring at LD2, we do not see them in our sparse dataset. Even with the data presented here, LD2 is obviously not like that of 29P/Schwassman-Wachmann 1, which has a radically more variable integrated brightness on these timescales \citep[see][]{2010MNRAS.409.1682T}.

\subsection{Dust Coma Characterization}
The visible images of LD2's dust coma allow characterization of its coma color, dust-production rates, and estimates of the dust outflow velocity. In this section we describe the methods used to estimate each property for LD2. Aperture photometry was used to measure the total apparent magnitude of LD2 based on the 2020 Jul. 3 Gemini GMOS images. Table \ref{tab:gemini_photometry} present these measurements for photometric aperture radii of: 0$''$.5, 5$''$, 10$''$, and 20$''$.

\begin{deluxetable*}{c|cccc}[ht!]
\tablecaption{Summary of UT Jul. 3 Gemini GMOS Photometry\label{tab:gemini_photometry}}
\tablenum{4}
\tablecolumns{5}
\tablewidth{0pt}
\tablehead{
\colhead{Filter} &
\colhead{$m(0''.5$)} &
\colhead{$m(5''$)} &
\colhead{$m(10''$)} &
\colhead{$m(20''$)}
}
\startdata
g$'$ & 20.16 $\pm$ 0.05 & 18.35 $\pm$ 0.05 & 18.08 $\pm$ 0.06  & 17.8 $\pm$ 0.1   \\
r$'$ & 19.47 $\pm$ 0.05 & 17.65 $\pm$ 0.05 & 17.34 $\pm$ 0.05  & 17.10 $\pm$ 0.07 \\
i$'$ & 19.15 $\pm$ 0.05 & 17.39 $\pm$ 0.05 & 17.06 $\pm$ 0.05  & 16.84 $\pm$ 0.08 \\
\enddata
\end{deluxetable*}
\FloatBarrier

\subsubsection{Coma Color}
Photometric measurements of LD2's coma were used to investigate the dust coma's color. We used the 5$''$ radius aperture photometry measurements for the color calculations for our cometary coma color measurements. LD2's coma on the date of Jul. 3 were typical of JFCs and other distantly active Centaurs \cite{}: $g' - r' = 0.70 \pm 0.07$ and $r' - i' = 0.26 \pm 0.07$.

\subsubsection{Coma Morphology}

\begin{figure}[ht!]
 \centering
\includegraphics[width=0.975\textwidth]{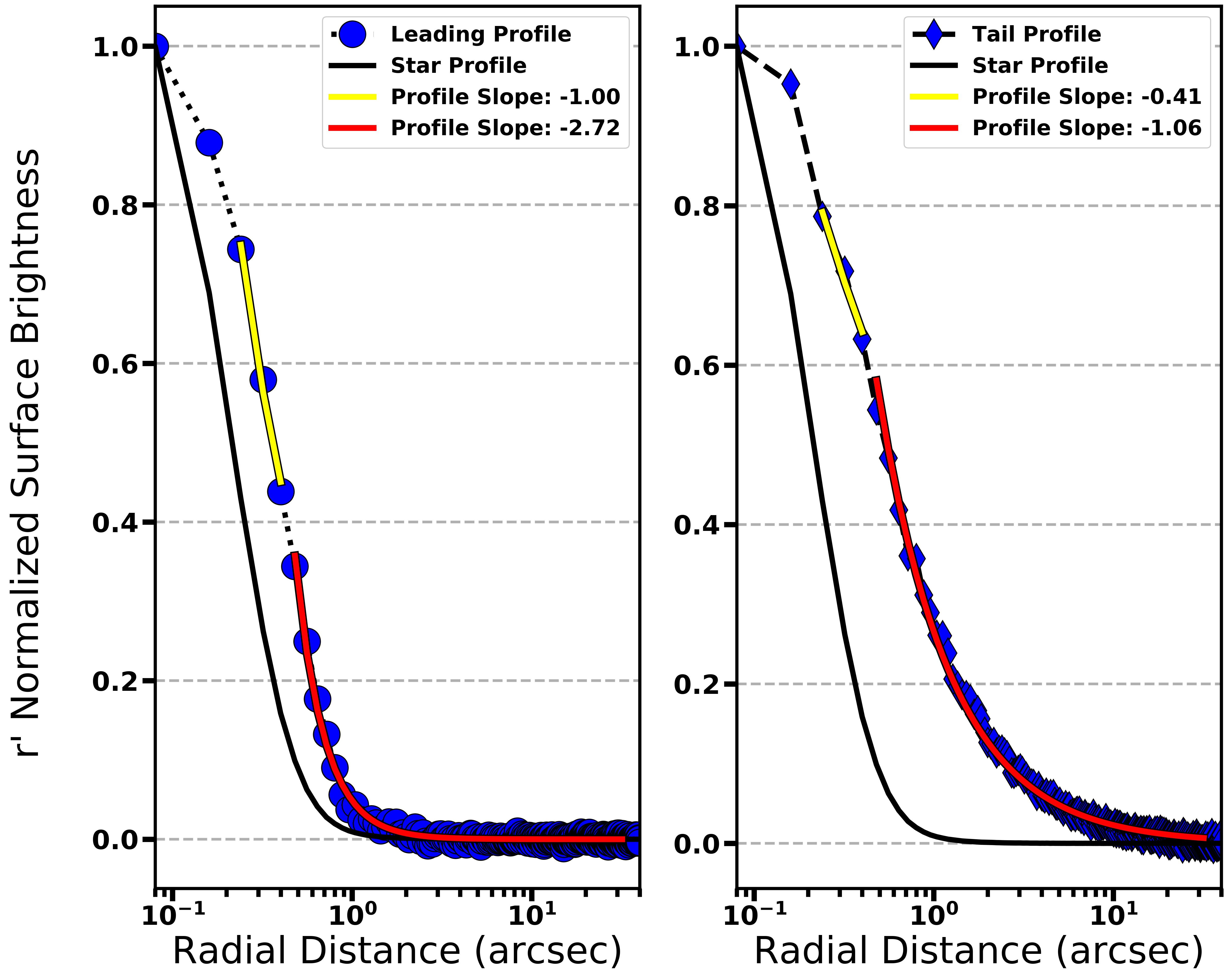}
\caption{Radial profiles of LD2 derived from the stacked GMOS r$'$ image. Coma profile fitted power law indices highlight the affects of solar radiation pressure. The projected cometocentric distance of 0$''$.5 for the leading profile was used as an estimated proxy for the coma dust grains' turn-back distance.}
\label{fig:gemini_r_profiles}
\end{figure}

LD2's coma morphology was investigated to estimate the dust coma's expansion velocity based on the turn-back distance detected in the Gemini imaging data. Due to the deeper coma surface brightness detection afforded by the stacked r$'$ image's equivalent 450 second exposure, we used this image for the estimation of the dust grain's outflow velocity. Figure \ref{fig:gemini_r_profiles} shows plots of the coma's radial profiles in the projected skyplane heliocentric velocity vector (velocity ``leading" coma direction) and negative of the heliocentric velocity vector (e.g., projected tail direction). For an isotropic emission of dust grains from the sunward facing side of LD2's nucleus, the coma's leading profile would have a power index $\sim$ -1 until the effects of solar radiation pressure turn-back the dust grains. \cite{mueller_2013} derives an equation for the turn-back distance due to solar radiation pressure, which we have inverted in order to estimate the dust coma's expansion velocity. The radial profile shown in Figure \ref{fig:gemini_r_profiles} for the leading direction becomes significantly steeper than a power index of -1 at an $\sim$ 0$''$.5 projected distance away from the nucleus's position. We use this distance as an estimate to the coma's projected turn-back distance: $d_{tb} = 2727$ km. Using the equation for turn-back distance \citep{mueller_2013}, we solve for the initial dust grain velocity's magnitude:
\begin{equation}
v = \Bigg[ \frac{2 d_{tb} \beta g \sin{\alpha} }{(\cos{\gamma})^2 } \Bigg]^{1/2},
\end{equation}

\noindent where $d_{tb}$ is the projected skyplane turn back distance of the dust grains, $\gamma$ is the angle between the initial direction of the dust grains and the skyplane, $\beta$ is the ratio of the radiation pressure acceleration to the acceleration due to solar gravity, $\alpha$ is the solar phase angle of the observations, and $g = G M_{\odot} / R_H^2$ is the solar gravitational acceleration on the dust grains ($G$ is the gravitational constant, $M_{\odot}$ is the Sun's mass and $R_H$ is the heliocentric distance of the dust grains). For LD2's $\sim$ 150 $\mu$m sized dust grains, as constrained by the contemporaneously acquired NIR spectral modeling (see Section \ref{sec:NIR_modeling} for further details and caveats), we use the equation from \cite{finson_probstein_1968} and \cite{burns_1979} that is applicable for dust grains larger than $\sim$ 1 micron to estimate a $\beta$ value for LD2:
\begin{equation}
\beta \approx \frac{4\times 10^{-7} \textrm{ m}}{d}
\end{equation} 

\noindent resulting in a estimated value of $\beta = 0.003$. The exact value for $\gamma$ of the dust grains most dominantly contributing to the detected coma profile's slope is unknown. Most probably it is the result of dust grains emitted over a continuum of angles. For this reason we calculate the outflow velocity for a range of plausible skyplane projected dust grain angles: $\gamma = 0^{\circ}$ ($v = 0.6$ m/s), $\gamma = 45.0^{\circ}$ ($v = 0.8$ m/s) and $\gamma = 80.0^{\circ}$ ($v = 3.3$ m/s).

\subsubsection{Dust Production Rate Estimates}
The $Af\rho$ parameter is a proxy for cometary activity first derived in \cite{ahearn_1984}

\begin{equation}\label{eq:Afrho}
	Af\rho(\textrm{cm}) = 10^{-0.4(m_c - m_{\odot})} \times  \frac{4 \Delta^2}{\rho} \bigg( \frac{R_H}{1 \textrm{ au}} \bigg)^2
\end{equation}

\noindent where: $\Delta$ is the geocentric distance of the comet in units of cm, $R_H$ is the heliocentric distance of the comet in units of au, $m_c$ is the measured magnitude of the comet within a circular aperture of projected radius $\rho$ (in units of cm), and $m_{\odot}$ is the solar magnitude of the Sun as measured in the same filter as $m_c$. For the Gemini GMOS r$'$ broadband image used for our $Af\rho$ calculations we use a value of $m_{\odot}$ = -27.04 from \cite{wilmer_2018}. The derivation of the $A f \rho$ parameter has many assumptions including: an isotropic emission of dust grains, a steady state dust production rate, no fragmentation of grains in the coma, and ignoring radiation pressure effects on dust grains. This ideal canonical coma will have a profile shape following a $1/\rho$ vs. cometocentric distance behavior. In practice the assumptions used to derive $A f \rho$ break down for real comae, but the calculation of $A f \rho$ provides a first order estimation of a nucleus's dust production rate if inspection of the coma's morphology is considered when choosing the photometric aperture radius used in Equation \ref{eq:Afrho}. Choosing a photometric aperture where the enclosed coma's profile more closely resembles a $1/\rho$ behavior provides a situation where the coma's morphology mimics the resulting behavior as the ideal coma used in the derivation of $A f \rho$. For the Jul. 3, 2020 Gemini r$'$ LD2 observations this photometric aperture radius was determined to be 0$''$.5 (see Figure \ref{fig:gemini_r_profiles}). Using the r$'$(0$''$.5) apparent magnitude measurement in Equation \ref{eq:Afrho} provides a value of $A f \rho$ = 484 $\pm$ 20 cm.

The derivation of $A f \rho$ provides an
expression relating it to the coma's dust production rate $\dot{M}_{dust}$ if one makes some highly-simplifying assumptions:

\begin{equation}\label{eq:mdot}
	\dot{M}_{dust} = \frac{8 a \rho_d v_d (Af\rho)}{3 A}
\end{equation}

\noindent where $a$ is the effective radius of the assumed-spherical typical dust grain, $\rho_d$ is the dust grain density, $v_d$ is the dust grain expansion velocity, and $A$ is the albedo of the dust grains. We note that while the conversion of $Af\rho$ to a true dust mass loss rate is fraught \citep{2012Icar..221..721F}, we aim here to only 
calculate a rough estimate.
We used the calculated $Af\rho(0''.5)$ value to estimate the dust production rate with Equation \ref{eq:mdot}. A value of $a$ = 75 $\mu$m was used based on the NIR spectral modeling. For the dust expansion velocity we used a value of $v_d$ = 0.8 m/s based on the $\gamma$ = 45.0$^{\circ}$ initial velocity direction. We chose limiting values of $A$ = 0.04-0.112 and calculated dust production rate estimates for assumed grains with bulk densities $\rho_d$ = 500 kg/m$^3$ and 1000 kg/m$^3$ found to be reflective of the range of values as observed by the Rosetta mission at 67P/Churyumov-Gerasimenko \citep{fulle_2016}. Using these assumptions we estimate dust production rates respectively of: 9.7 $\pm$ 0.4 kg/s and 19.4 $\pm$ 0.8 kg/s for a value of $A$ = 0.04 and 3.5 $\pm$ 0.1 kg/s and 6.9 $\pm$ 0.3 kg/s for a value of $A$ = 0.112.

In \citep{Lowry1999}, $Af\rho$ values are provided for several Jupiter Family comets beyond 3 AU, which range from 7.4 cm to 228.8 cm. By comparison, LD2 shows a much higher $Af\rho$ of 484 cm, which is supported by the well defined dust tail in \ref{fig:gemini_images}. In terms of $Af\rho$, LD2 thus appears more active than a JFC at a comparable distance.

\subsection{Upper Limit to CO Production and source of activity}
\label{sec:COupperlimit}

LD2 has sustained a dust coma for many months beyond $\sim$ 4 au, which is too far from the Sun for water-ice sublimation, the most common trigger of activity for comets close to the Sun, to proceed efficiently. Carbon monoxide emission is observed in 5 - 20 \% of the gas comae of comets and it readily outgases and sublimates well beyond 4 au. Thus, it was a prime suspect for searching in LD2. 
We did not detect a CO line in the SMT data and correspondingly we derive a 3-$\sigma$ upper limit to the line area  of 0.04 K kms$^{-1}$ from ${3{\textit{rms}\over\eta_b}\sqrt{\Delta v_c\Delta v_{line}}} $, where \textit{rms} = 0.012 K, beam efficiency, $\eta_b$, = 0.74, $\Delta v_c$ is the velocity width of the channel of the CO 2-1 line, 0.33 km s$^{-1}$,  and an line integration width of $\Delta v_{line}$ = 2.0 km s$^{-1}$, consistent with the assumed expansion velocity of $v_{exp}$ = 0.5 km s$^{-1}$ \citep[see][]{Mangum2015}. Using 0.5 km s$^{-1}$ for the expansion velocity is at the upper end of the expected range upper end for an object at such a large heliocentric distance and leads to the strictest upper limit for CO \cite{Gunnarsson03, wierzchos2018}. We modeled the gas emission using this line area upper limit, assuming an optically thin gas, a kinetic temperature of 20K \citep{biver02},  and using  non-LTE molecular excitation calculations, including contributions from collisions with H$_2$O, fluorescence, and cosmic background radiation \citep[see][]{wierzchos2018}, to derive a 3-$\sigma$ upper limit of $Q(CO) <$ $4.4\times10^{27}$ mol s$^{-1}$ ($<$ 205 kg s$^{-1}$). This model also assumes isotropic outgassing of the CO from the nucleus and a photodissociation decay model \citep{Haser57} with photodissociation rate of 1.33 x 10$^6$ s at 1 au for a quiet Sun \citep{Huebner1992}. Radiative decay from the J=2 level occurs with increasing distance from the nucleus and thus the production rate upper limit may be slightly underestimated \citep{GarciaBerrios2020}.

These calculations represent an upper limit to the amount of CO arising from a nuclear source. CO emission could also arise from photodissociation of other volatiles in the coma, such as H$_2$CO, CH$_3$OH, or CO$_2$ \citep{Pierce2010}. CO emission may also arise from icy grains in comets \citep{gunnarsson08, Womack2017DistantReview}. The ratio of native to distributed CO production rates in a sample of several comets was determined to be approximately 80\% to 20\%, respectively \citep{Disanti03}. Given the lack of detection of CO or any other molecule, it is beyond the scope of this paper to speculate further on native vs. distributed contributions to CO.

Our CO upper limit of $Q(CO) < 4.4\times 10^{27}$ mol s$^{-1}$ is substantially lower than what is typically observed for the always active centaur 29P/Schwassmann-Wachmann 1 at $\sim$ 6 au, ($Q(CO) \sim 1-3 x 10^{28}$mol s$^{-1}$, and the great long-period comet C/1995 O1 (Hale-Bopp) at 4.6 au ($Q(CO) \sim  2 x 10^{28}$mol s$^{-1}$. However, we urge caution about making such direct comparisons of production rates without considering distance from the Sun, or size, chemical composition or physical construction of the nucleus. For example, Hale-Bopp and 29P have nuclei $D \sim$ 60 km in diameter, much larger than that of LD2. In lieu of a detailed model of energy balance and chemical composition of the nucleus, we can still tell a great deal by comparing the gas production rates scaled for nucleus surface area, which we call ``specific production rates,'' $Q(CO)/D^2$.  If we assume that LD2's diameter is $D \sim$ 2 km (consistent with this paper's upper limit), then we would get a specific production rate of $Q(CO)/D^2$ $<$ 10$^{27}$ mol s$^{-1}$ km$^{-1}$. This is much higher than the CO specific production rates for Hale-Bopp of $\sim$ 6 x 10$^{24}$ mol s$^{-1}$ km$^{-1}$ at 4.6 au (the same distance as LD2) or 29P of $\sim$ 8 x 10$^{24}$ mol s$^{-1}$ km$^{-1}$ at 6 au. Our upper limit to LD2's specific production rate places it well above the line dividing CO-depleted from CO-rich gas comae for 29P, Hale-Bopp, Echeclus, Chiron and other Centaurs, and thus our value does not distinguish it from either case (see Figure 3 in \citet{wierzchosecheclus2017}.) If LD2's diameter is $D \sim$ 2 km, then its production rate would be $Q(CO) \sim 3x 10^{25}$ mol s$^{-1}$ if it had a similar specific production rate to Hale-Bopp or 29P. An upper limit comparable to this value or lower would be needed to test models, with correspondingly lower values needed if the nucleus were smaller. Future work to directly measure or better constrain LD2's CO production rate to values closer will be critical in making comparisons to 29P, Echeclus, Hale-Bopp, or other distantly active objects, and thus understanding which volatiles might be driving its activity more definitively.

Thus, our observations cannot put constraints on whether CO is present in large enough amounts to be involved in producing the observed dust coma.\footnote{Spitzer measurements of LD2 were used to derive a comparable 3-$\sigma$ upper limit of 4.8x10$^{27}$ mol s$^{-1}$ for CO, assuming that all gaseous emission in the 4.5 $\mu$m band was from  CO (both CO and CO$_2$ emit in that bandpass) \citep{Bolin2020}}.  For reasons described above, water-ice sublimation is not likely to play a large role in LD2's activity over the last year or two, although it will become important as it moves inward (assuming it is present in the nucleus). CO emission is still possible, as are other candidate volatiles, such as CO$_2$, CH$_4$, O$_2$, and even N$_2$, which are much. more difficult to measure with current telescopes and instrumentation \citep{Womack2017DistantReview}.

\subsection{Dust and (Possible) Ice}
\begin{figure}[ht!]
 \centering
\includegraphics[width=0.975\textwidth]{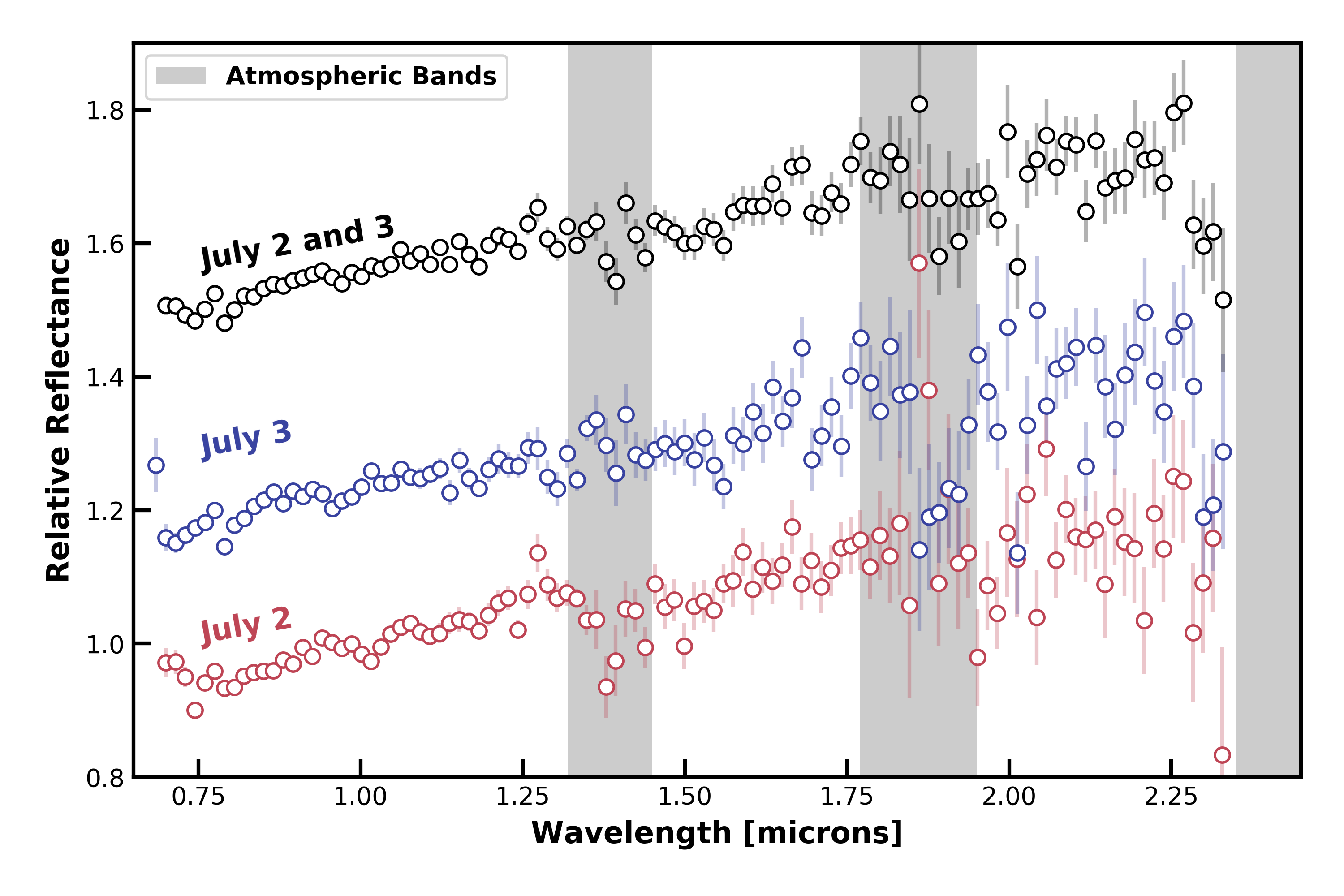}
\caption{The reflectance spectra of the inner coma (extraction aperture radius of $2.0\arcsec$) of P/2019 LD2 on July 2 (red), July 3 (blue), and the combination of both dates (black). All spectra were normalized at 1.5 microns, offset for clarity, and combined at a spectral resolution of $R\sim100$. The grey shaded regions are areas of higher telluric absorption, and thus lower signal to noise ratios.}
\label{fig:irtf_1}
\end{figure}

The retrieved infrared reflectance spectra are shown in Figure \ref{fig:irtf_1}. The spectra from July 2 and 3 are consistent in spectral slope, showing a gentle overall red (increasing reflectivity as a function of wavelength) slope. Using the $S'$ framework \citep{1990Icar...86...69L}, the average slope of the two is $S'=2.0\pm0.2$ (in units of $\%$ per $0.1\mu{m}$) over the wavelength range $0.8-1.25 \mu{m}$. The extracted spectrum was not statistically different using smaller or larger apertures, which is different from the spectral behavior of 174P/Echeclus as observed with the same instrument while active \citep{2019AJ....158..255K}. The spectra were combined at an $R\sim100$ to properly boot-strap the errors from the data, as spextool \citep{2004PASP..116..362C} can underestimate errors in certain circumstances. The reflectivity near $\sim1.5\mu{m}$ appears slightly curved or depressed, there is a less obvious drop in reflectivity near $\sim1.95-2.0\mu{m}$, and the flux appears to drop precipitously beyond $\sim2.3\mu{m}$. The former two features appear plausibly real with apparent absorption depths near $\sim5$\%, while the latter likely results from decreasing signal-to-noise at the longest wavelengths due to rapidly increasing atmospheric absorption. We further note for clarity that the proximity of the proposed $\sim1.95-2.0\mu{m}$ drop in reflectivity to the strong telluric absorption band complicates any quantitative analysis of the feature (depth, center) as the shorter-wavelength continuum is challenging to discern without higher quality data. A similar decrease in flux near $\sim1.5\mu{m}$ and $\sim2.0\mu{m}$, caveats accepted, could be related to the presence of water ice (either pure or incorporated into mixed grains) in the coma of LD2. The origin of the spectral continuum and features superimposed on it are modeled using a variety of techniques in Section~\ref{sec:NIR_modeling} to constrain the composition of ejected material in the coma of LD2.

\subsection{Near-Infrared Spectral Modeling}\label{sec:NIR_modeling}
Modeling the reflected light from the solid coma of an active object requires a choice about what is the correct physical model to apply -- if the grains are large, then a Hapke-like model can be applied (\citealt{1993tres.book.....H}, see also \citealt{2007Icar..190..284S, 2009AJ....137.4538Y,2014Icar..238..191P}), while if the grains are similar in size or smaller than the wavelength, then a model based on Mie scattering should be applied \citep{2014ApJ...784L..23Y, 2018ApJ...862L..16P}. In the absence of additional information to constrain the problem (e.g. longer wavelength observations of thermal emission, measurements over a range of heliocentric distances), we blindly employ each kind of model with a subset of optical parameters and constituent species to constrain what the coma \textit{might} contain to guide later analysis. The constituent species of this compositional modeling are carbon (the optical constants of amorphous ``glassy" carbon of \citealt{edoh_optical_1983} and the amorphous carbon of \citealt{1991ApJ...377..526R}) and water ice (both amorphous and crystalline at various temperatures from \citealt{2008Icar..197..307M}). In general, the carbon acts as a featureless, dark, and red-sloped absorbing material which counterbalances the bright steeply blue-sloped spectrum of water ice. The two samples of amorphous carbon are redder and bluer than the overall continuum of our spectrum of LD2 (though both still are red), providing end-members over which to impose the water ice spectral features should they be found to be significant in the modeling. For both the large grain (Hapke) and small grain (Mie) cases, we consider a situation where the icy grains and dust grains are distinct (``linear mixing"), and for the large grains we also consider a case where a single dust grain can contain both ice and dust (``intimate mixing"). We thus have three ``classes" of models with several combination of optical constants to test against the data. As an additional test, we also employ a large grain (Hapke) case where the two constituent inputs are the two kinds of carbon employed. In general, a ``good fit" should both have a reasonable reduced $\chi^2$ value near $\sim1.0$ without over-fitting or over-interpreting the data.

The two Hapke models (linear and intimate mixing), the single Hapke/Mie model, and the no-ice all carbon model are presented in Figure 5. The carbon and ice components were forced to have the same grain size in the linear cases, and in all cases the grain size and mixing ratio (either by area or mass) were the two fit parameters. Reported values are for crystalline ice at 130 K, the amorphous carbon of \citealt{edoh_optical_1983}, and all $\chi^2$ values are reduced . The two ice-and-carbon Hapke models as shown have similar reduced chi-square values of $\chi^2 = 1.07$ (linear) and $\chi^2 = 1.16$ (intimate). In general, the intimate mixing model contains more ice ($9\pm2$ weight \%), while the linear mixing model contains much less ice ($2 \pm 0.6$ areal \%). Notably, the linear mixing model converges on a grain size that is at the limit of what the Hapke formalism technically allows ($\sim2.28\mu{m}$), but the model shown is for a compliant grain size ($\sim2.5\mu{m}$) with an identical reduced $\chi^2$ value. The intimate mixing model weakly prefers a large grain size around $\sim150\mu{m}$. The no-ice model is a worse fit to the data ($\chi^2 = 1.57$), as it can match the overall slope well (as expected), it performs much worse after $\sim1.5\mu{m}$ by attempting to go through the area of lower reflectance. The Mie-scattering model (which assumes fine grained pure ice and larger-grained dust) has $\chi^2 = 1.10$, an areal ice fraction of $1.2 \pm 0.1$ areal \%, and an effective dust size near $6-7\mu{m}$. Somewhat notably, the $\chi^2$ value drops slightly for the Linear mixing and Mie models if amorphous ice is used, while similar qualities of fits were retrieved with all temperatures and phases of ice for the Hapke models. Amorphous ice, if present, is not stable for long time periods even at LD2's perihelion distance, suggesting it would have been released recently from the nucleus. While this small change is not adequate to provide strong evidence for amorphous over crystalline ice, it does better fit the shape and center of dip in reflectance near $\sim 2.0\mu{m}$. Thus, if the ice were amorphous, it might be a better estimate of the overall ice content of the upper regolith of the Centaur, while crystalline ice is stable indefinitely and thus could "build up" in the coma as activity persisted. Further observations can thus constrain the ice content of the outermost layers of the nucleus, given higher signal to noise. We note that these results did not statistically change with different binning resolutions or normalization wavelengths, and while the quoted fit parameters are from fits where areas of high telluric absorption were not included, the retrieved fit parameters did not change significantly (the $\chi^2$ values are simply higher). The quality of the fits for all models considered is poorer at long wavelengths, suggesting that most of the constraining power is in the region surrounding the $\sim1.5\mu{m}$ feature, but not including this region makes the continuum fit beyond $\sim2.0\mu{m}$ much worse. The difference between the ice and no-ice models is most stark when the wavelength range is constrained to be immediately surrounding the $1.5\mu{m}$ feature (often producing a multiple-sigma difference between the data and model in that region) and becomes less prominent as more data is added in allowing the all-continuum no-ice model to slowly improve its reduced $\chi^2$ value. To be clear, for all normalization values and wavelength ranges tested, the ice-containing models outperformed the no-ice model (indeed, the linear mixing ice-and-carbon model could choose to converge on the no-ice model exactly) but the specific ratio of qualities of fits varied with different ice or carbon choices as expected. In summary, all three models that allowed for it are significantly improved with the addition of water ice at the $\sim$ few \% level, but the quality of the data prevents a reliable determination of its crystallinity or total abundance. While an all-dust no-ice model is not truly ruled out by the data presented, we believe the detection of ice at the $\sim$ few $\%$ level as the fits were both quantitatively better and qualitatively better matched the observed feature(s) in the data. Higher SNR or higher resolution observations in the future will be critical to better understanding the properties and abundance of ice in the solid coma of LD2 and how it evolves in time.

\begin{figure}[ht!]
 \centering
\includegraphics[width=0.975\textwidth]{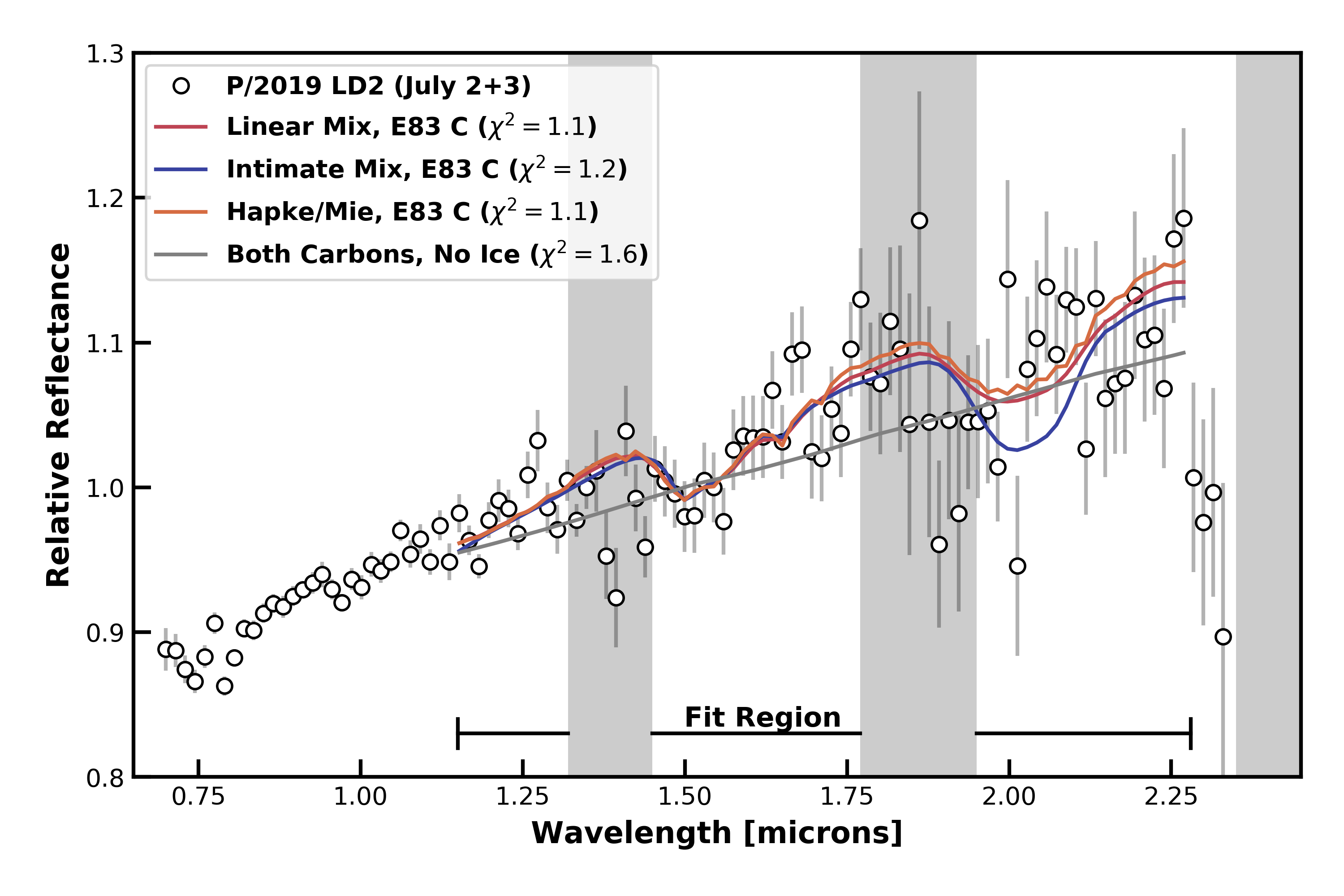}
\caption{The reflectance spectra  of P/2019 LD2 \label{fig:general} on July 2 and July 3 (black) compared with the best-fit large grain linear mixing model (red), intimate mixing model (blue), and Hapke (dust) and Mie (ice) model (orange) of crystalline water ice at 130 K and different kinds of amorphous carbon, as well as model that employs both amorphous carbons and no ice (grey), all normalized at $1.5\mu{m}$. Refer to the text for details of the models.}
\end{figure}

\section{Discussion and Conclusions}
\subsection{LD2 as a Future JFC}

Is P/2019 LD2 (ATLAS) what we expect a soon-to-be JFC to be like? The comparative importance of the object is tied to whether it can be directly related to its inner Solar System relatives. LD2's current properties need to be understood within the context of other active Centaurs (especially Gateway objects; \citealt{Steckloff:2020}) and conventional JFCs if we are to properly use it to understand the Centaur-JFC transition. Is it `typical' or is it anomalous somehow? When it enters the inner Solar System in 2063, will the scientific community be watching a normal Centaur become a normal JFC, or will different processes be in play? In this work, we have made some of the first quantitative measurements of the properties of LD2 and laid some groundwork for future observations. 

Based on LD2's brightness in the most conclusive March 2017 DECam precovery images, we estimate that LD2's nucleus is less than $\sim 1.2$ km in radius, which is far less than the other best-studied Centaurs like 29P ($\sim30$ km, \citealt{schambeau_2015}), Chiron ($\sim70$ - 109 km, \citealt{2004A&A...413.1163G, fornasier13}), or Chariklo ($\sim125$ km, \citealt{2014Natur.508...72B}). This is, however, within the range of common sizes for JFCs \citep{2004Icar..170..463M, 2017AJ....154...53B}. While this size difference is not a surprise on a number frequency basis, small Centaurs of this size are still rare and challenging to find. No outbursts were detected in our archival Catalina Sky Survey data, and the brightness of the object only changed gradually with time. The brightness of LD2 during the Gemini, IRTF, and SMT observations was stable on days-long timescales. In general, this is unlike the extremely outburst prone Centaur 29P/S-W 1 \citep{whipple_1980, 2010MNRAS.409.1682T, miles_2016}.

LD2's activity similarly is plausibly typical. We used our millimeter-wavelength observations of the J=2-1 rotational transition of carbon monoxide (CO) emission to calculate a production upper limit of $Q(CO) < 4.4\times10^{27}$ mol s$^{-1}$  when it was $r_H$ = 4.58 au from the Sun. CO emission has not been detected in many comets beyond 4 au, but for comparison, CO emission was detected in the great, long-period comet C/1995 O1 (Hale-Bopp) with Q(CO) $\sim$ 2 x 10$^{28}$ mol s$^{-1}$ at 4.6 au \citep{biver02,Womack2017DistantReview}, repeatedly seen in the Centaur 29P/Schwassmann-Wachmann 1 at $\sim$ 6 au with Q(CO) $\sim$ 3 x 10$^{28}$ mol s$^{-1}$ and marginally detected in the Centaur 60558 (Echeclus) at 6.1 au with Q(CO) $\sim$ 8 x 10$^{26}$ mol s$^{-1}$ \citep{wierzchosecheclus2017}. However, as discussed in Sec. \ref{sec:COupperlimit}, nucleus size cannot be ignored when comparing production rates: these objects are much larger. If LD2's nucleus was sufficiently radically larger ($r\sim 20 km$), then our CO upper limit might make it appear CO-depleted compared to a Hale-Bopp or 29P/S-W 1 \citep{wierzchosecheclus2017}, but for a $\sim$ km-scale nucleus this is a highly plausible upper limit. In other words, the small (typical) size of LD2 combined with a lack of any detectable outburst makes our non-detection of CO unsurprising. 

The spatial profiles from visible and near-infrared observations support ongoing activity since at least 2018. LD2 stands in stark contrast to the more explosive and stochastic outburst-centric activity of centaurs like Echeclus \citep{2019AJ....158..255K} or 29P. The strong but ongoing activity of LD2 is well within the expectation of a JFC-like object that has yet to spend significant time within the inner Solar System. Its activity pattern is not quite like that of the ultra-active 29P/S-W 1 or the large outbursts of Echeclus with little activity in between, but it is also not so depleted in substances volatile at $\sim 5$ AU that its activity in that region is minimal. LD2's $Af\rho$ presented here is higher than that of a typical JFC at comparable distance from the Sun, which corresponds to a typical JFC's aphelion. For example \citet{2005MNRAS.358..641L}, \citet{2008MNRAS.385..737S}, and \citet{epifani_2009} report observations of numerous distant JFCs, most of which are consistent with $Af\rho$ of either zero or only tens of cm. A comparison of LD2's dust coma activity with other JFCs also having a large perihelion is poorly understood at this time. LD2 dust activity is higher when compared to JFCs that have perihelion less than $\sim$4 au, but we do not know at this time how it compares to other JFCs with $q >$ $\sim$ 4.5 au. While there have been many recent discoveries over the past decade of JFCs with perihelion near Jupiter, this population's long-term dust comae activity behaviors and ensemble properties have been poorly studied or at least as reported in the literature. Several groups have reported snapshot dust coma characterization for several active Centaurs (\citealt{jewitt_2009, wong_2020, epifani_2017}) and high perihelion JFCs (e.g., JFCs with $q >$ 4.5 au; \citealt{schambeau_2018}). The many recent and ongoing surveys have discovered $\sim$ 40 such JFCs with large perihelia. As a community we should also focus attention on understanding this populations ensemble properties and placing LD2's activity behaviors as it transitions from Centaur to JFC in context with the wider population.

The very probable detection of water ice at the $\sim$ few \% level in the inner coma of LD2 also bolsters the claim that the object is both `typical' and has yet to spend much time in the inner Solar System. While a typical JFC rarely has water ice detected within its coma, the lifetimes of icy grains is dramatically shorter at smaller heliocentric distances (see, e.g., \citealt{2018ApJ...862L..16P}). The most hyper active comets in the inner Solar System become detectably icy when their comae ice fractions approach the $\sim$ 10s of \% level \citep{2014ApJ...784L..23Y, 2014Icar..238..191P, 2018ApJ...862L..16P}, more than detected here. However, the detection of water ice within 9P/Tempel 1 \citep{2007Icar..190..284S} strongly suggests that small icy grains are a component of even non-hyperactive nuclei interiors near the surface. Thus, LD2's barely icy coma would be extremely challenging to detect in the inner Solar System, but consistent with that of a typical nucleus that has such grains available. If icy grains were detected during an outburst but \textit{not} during transient activity, that might imply that the icy grains are buried at some depth and are not available for traditional mass loss in the upper regolith. The detection of ice in the case of ongoing activity suggests that icy grains are at least mixed into the uppermost regolith at some level, which would not expected if the surface had been heated beyond where water ice had been stable. It is unclear whether this fractional icy component of the solid coma would have been detected if observations occurred when the object first activated, but the current observational understanding provides evidence that at least the few topmost layers of the regolith now have not been heated to the point of desiccation. Comet 103P/Hartley 2 was observed in situ to have significant amounts of water ice detected in its coma \citep{2014Icar..238..191P}, likely driven off the surface by a more volatile species. Indeed, this behavior is indicative of exposures of ice more volatile than crystalline water \citep{Steckloff:2016,Steckloff:2018}, consistent with the dynamical interpretation that LD2 has spent little time in the inner solar system \citep{Steckloff:2020}.

The observations and derived conclusions that we present in this paper are consistent with the hypothesis that LD2 is very likely as a typical \textit{small} Centaur on its way to becoming a typical JFC. However, the uniqueness and importance of studying LD2 stem from it being an object that has likely never entered the inner Solar System for any significant period of time. It may be one of the best examples of a \textit{transient} object caught in the act of transition, both dynamically and physically.

Following this conjecture, ongoing studies of LD2 at the present date and over the next few decades could greatly advance our understanding of the life cycle of Jupiter Family Comets, as well as the properties of the smallest and most numerous Centaurs, as well as enhance our understanding of recent entrants to the Jupiter Family.

\subsection{Future Observational Priorities}
The present study leaves open several key questions which could guide the next steps of observations to help characterize this intriguing object. The primary goal over the next few decades should be to simply observe the object frequently to monitor its activity and thereby look for changes. At $m_v \sim 18-19$, the object is observable by many non-professional astronomers and will likely be detected by the major asteroid surveys (ATLAS, Catalina Sky Survey, PanSTARRS, etc) which regularly survey to two magnitudes deeper. This will also be invaluable to scheduling and planning more detailed follow up work with larger telescopes, especially after the object moves into a more distant (and thus fainter observationally) orbit in 2028. (Its perihelion distance will be beyond 5 AU from the Sun, which is further than activity has been detected in its current orbit, suggesting that the object will be both intrinsically fainter and further.) As the object moves away from perihelion in the coming years, it will likely move beyond the reach of amateurs and surveys (with the exception of the Rubin Observatory's LSST) prior to that date. We are particularly interested in continued monitoring of the ice content and outgassing patterns (including searches for gas species), but all techniques and observations will be useful. Systematic programs to monitor the activity of JFCs with perihelia at similar heliocentric distances will also be key to contextualizing LD2's activity and properties \citep{schambeau_2018}.

To facilitate and encourage further observations of P/2019 LD2 (ATLAS), several of the authors of this manuscript have organized an observing campaign website to plan and coordinate observations of all kinds (\url{https://observe-ld2.blogspot.com/}). Interested observers can submit information about their ongoing and planned observations and see contact information for the campaign organizers and other observers.

\section{Summary}

In this paper, we have detailed contemporaneous observations at visible, near-infrared, and millimeter wavelengths as well as visible-wavelength precovery observations of the active Centaur P/2019 LD2 (ATLAS). LD2 is currently in a Gateway orbit \citep{Sarid:2019} and will very likely become a  Jupiter Family of Comet (JFC) in 2063 \citep{Kareta:2020, 2020arXiv200713945H}. This is very likely the first time that LD2 has made this transition \citep{Steckloff:2020}, meaning that this is the first recognized time that the community can observe an object move from one population to the other. This transition is marked by changes in sublimating volatiles and activity character and is in general quite poorly understood \citep[see][]{Womack2017DistantReview}. If put into the proper observational context, LD2 could thus be an extremely useful case study to understand how this transition affects these objects and thus better understand the evolutionary history of the Centaurs and JFCs in general. 

Precovery observations from the Dark Energy Survey's \citep{abbott_2018} DECam instrument \citep{decam_2015_aj} from 6 March 2017 find no detectable object in the vicinity of where LD2 should be down to $r'\sim23.8$, which suggests the radius of LD2 to be $\sim1.2$ km or smaller given a 5$\%$ visible albedo. Observations by DECam in August 2018 show the object to be somewhat extended and intrinsically brighter, suggesting that activity had started in earnest by that point. Archival detections by the Catalina Sky Survey are marginal in May and June 2018 (it is not possible to tell if the object was active or not), but since April 2020 the object has been sampled and detected well by the survey. The sparse photometry shows only gradual changes in brightness with no obvious signs of any abrupt brightening (outburst events.) During the July 2020 observations detailed below and in Section 2, the activity of LD2 appeared stable.

Observations in visible broadband colors at Gemini North on 3 July 2020 show spatial profiles consistent with ongoing and stable activity and consistent colors at multiple aperture sizes. The measured coma colors are $g' - r' = 0.70 \pm 0.07$ and $r' - i' = 0.26 \pm 0.07$ and the inferred dust production rate based on the measured $Af\rho$ is $\sim 10-20$ kg/s (assuming various grain properties). The $Af\rho$ itself is far higher than that of typical JFCs at this heliocentric distance, which corresponds to a JFC's typical aphelion. We used LD2's coma morphology as seen in the $r'$ filter to estimate the dust coma's outflow velocity between $ v \sim 0.6-3.3$ m/s. The reflectance spectra obtained of LD2's inner coma in the NIR from the NASA Infrared Telescope Facility (IRTF) on 2-3 July 2020 are red and show evidence for weak absorption features at $1.5$ and $2.0\mu{m}$, which we interpret as a small amount of water ice in the coma of LD2. Depending on spectral modeling parameters, we find values ranging from approximately 1\% (for linear mixing) to 11\% (for intimate mixing), significantly less than for hyperactive JFCs (tens of \%). Models with large grains (Hapke) and small grains (Mie scattering) appear to fit the data similarly well, but higher quality data would likely be able to distinguish the ice properties (crystallinity, mixing method, etc.) in the future.

Two nights (2020 July 2-3) of integration towards LD2 at the Arizona Radio Observatory Sub-Millimeter Telescope (SMT) did not yield a CO detection, and we used the measurements to derive CO production rate upper limit of $Q(CO) <  4.4\times10^{27}$ mol s$^{-1}$ (3-$\sigma$). By comparing specific production rates of LD2 with other Centaurs and the great comet Hale-Bopp we conclude that this is not a significant upper limit and that CO may still play a large role in the activity. If LD2's behavior is similar to that of the distantly active 29P and Hale-Bopp, then we predict its CO production rate will be approximately $Q(CO) < 3x 10^{25} mol s^{-1}$.

The size of LD2, the character of its activity, the colors of its coma, its lack of tremendous CO production, and the small amount of water ice in its inner coma are all consistent with the idea that LD2 is a typical Centaur (though still smaller than the best-studied Centaurs) that will become a typical Jupiter Family Comet. Understanding LD2's modern properties and how they evolve over the next few decades will be critical to understanding the physical differences between the Centaurs and the JFCs and how the transition between populations affects the individual small bodies. In Section 4, we describe our proposal for a long baseline observational campaign of LD2 to attempt to learn as much as we can about this important object and thus better understand the modern properties and evolutionary history of the Centaurs and the JFCs.

\section{acknowledgments}
The authors are extremely thankful to the Directors and staff at Gemini North, the NASA IRTF, and the ARO SMT for awarding us Director's Discretionary Time, especially on such a tight schedule and with such a quick turnaround given the nature of the project. This project would not have been possible without their hard work, especially during the pandemic. Our hearts go out to all those affected by these trying times, and we are extremely grateful to be able to pursue this work, now more than ever. We would also like to thank Benjamin N.L. Sharkey for useful discussions about spectral modeling. This material is based in part on work done by M.W. while serving at the National Science Foundation. The SMT is operated by the ARO, the Steward Observatory, and the University of Arizona, with support through the NSF University Radio Observatories program grant AST-1140030.

The authors wish to recognize and acknowledge the very significant cultural role and reverence that the summit of Mauna Kea has always had within the indigenous Hawaiian community. We are most fortunate to have the opportunity to conduct observations from this mountain. The IRTF is operated by the University of Hawaii under Cooperative Agreement no. NCC 5-538 with the National Aeronautics and Space Administration, Office of Space Science, Planetary Astronomy Program. Based on observations obtained at the international Gemini Observatory, a program of NSF’s NOIRLab, which is managed by the Association of Universities for Research in Astronomy (AURA) under a cooperative agreement with the National Science Foundation. on behalf of the Gemini Observatory partnership: the National Science Foundation (United States), National Research Council (Canada), Agencia Nacional de Investigaci\'{o}n y Desarrollo (Chile), Ministerio de Ciencia, Tecnolog\'{i}a e Innovaci\'{o}n (Argentina), Minist\'{e}rio da Ci\^{e}ncia, Tecnologia, Inova\c{c}\~{o}es e Comunica\c{c}\~{o}es (Brazil), and Korea Astronomy and Space Science Institute (Republic of Korea). The SMT is operated by the Arizona Radio Observatory (ARO), Steward Observatory, University of Arizona.

This project used public archival data from the Dark Energy Survey (DES). Funding for the DES Projects has been provided by the U.S. Department of Energy, the U.S. National Science Foundation, the Ministry of Science and Education of Spain, the Science and Technology FacilitiesCouncil of the United Kingdom, the Higher Education Funding Council for England, the National Center for Supercomputing Applications at the University of Illinois at Urbana-Champaign, the Kavli Institute of Cosmological Physics at the University of Chicago, the Center for Cosmology and Astro-Particle Physics at the Ohio State University, the Mitchell Institute for Fundamental Physics and Astronomy at Texas A\&M University, Financiadora de Estudos e Projetos, Funda{\c c}{\~a}o Carlos Chagas Filho de Amparo {\`a} Pesquisa do Estado do Rio de Janeiro, Conselho Nacional de Desenvolvimento Cient{\'i}fico e Tecnol{\'o}gico and the Minist{\'e}rio da Ci{\^e}ncia, Tecnologia e Inova{\c c}{\~a}o, the Deutsche Forschungsgemeinschaft, and the Collaborating Institutions in the Dark Energy Survey.
The Collaborating Institutions are Argonne National Laboratory, the University of California at Santa Cruz, the University of Cambridge, Centro de Investigaciones Energ{\'e}ticas, Medioambientales y Tecnol{\'o}gicas-Madrid, the University of Chicago, University College London, the DES-Brazil Consortium, the University of Edinburgh, the Eidgen{\"o}ssische Technische Hochschule (ETH) Z{\"u}rich,  Fermi National Accelerator Laboratory, the University of Illinois at Urbana-Champaign, the Institut de Ci{\`e}ncies de l'Espai (IEEC/CSIC), the Institut de F{\'i}sica d'Altes Energies, Lawrence Berkeley National Laboratory, the Ludwig-Maximilians Universit{\"a}t M{\"u}nchen and the associated Excellence Cluster Universe, the University of Michigan, the National Optical Astronomy Observatory, the University of Nottingham, The Ohio State University, the OzDES Membership Consortium, the University of Pennsylvania, the University of Portsmouth, SLAC National Accelerator Laboratory, Stanford University, the University of Sussex, and Texas A\&M University.
Based in part on observations at Cerro Tololo Inter-American Observatory, National Optical Astronomy Observatory, which is operated by the Association of Universities for Research in Astronomy (AURA) under a cooperative agreement with the National Science Foundation.

Based on observations obtained at the international Gemini Observatory (Program IDs: GN-2020A-DD-202 and GN-2020B-Q-405), a program of NSF’s NOIRLab [processed using the Gemini IRAF package], which is managed by the Association of Universities for Research in Astronomy (AURA) under a cooperative agreement with the National Science Foundation. on behalf of the Gemini Observatory partnership: the National Science Foundation (United States), National Research Council (Canada), Agencia Nacional de Investigaci\'{o}n y Desarrollo (Chile), Ministerio de Ciencia, Tecnolog\'{i}a e Innovaci\'{o}n (Argentina), Minist\'{e}rio da Ci\^{e}ncia, Tecnologia, Inova\c{c}\~{o}es e Comunica\c{c}\~{o}es (Brazil), and Korea Astronomy and Space Science Institute (Republic of Korea).

This research made use of ccdproc, an Astropy package for image reduction \citep{matt_craig_2017}.

\vspace{5mm}
\facilities{Gemini-N (GMOS-N), IRTF (SpeX, MORIS), ARO:SMT}

\software{NumPy \citep{harris_array_2020}, SciPy \citep{virtanen_scipy_2020}, AstroPy \citep{astropy_collaboration_astropy_2013,astropy_collaboration_astropy_2018}, Gemini IRAF \citep{gemini_iraf_2016}, spextool  \citep{2004PASP..116..362C}}


\end{document}